\documentclass[preprint,12pt]{article}

\usepackage{graphicx}  
\usepackage{amsmath}   
\usepackage{amssymb}   

\usepackage[active]{srcltx}
\usepackage[compress]{cite}

\usepackage{soul}

\def\eq#1{{Eq.~(\ref{#1})}}

\def\cc{{cosmological\ constant}}
\newcommand{\LL}{Lanczos-Lovelock}

\title{Gravity \st{and} is Thermodynamics}
\author{T. Padmanabhan\\
  IUCAA, Pune University Campus,\\
  Ganeshkhind, Pune- 411 007.\\
  {\small {email: paddy@iucaa.ernet.in}}}

\begin{document}

\maketitle

\begin{abstract}
The equations of motion describing all physical systems, excluding gravity,  remain invariant if a constant is added to the Lagrangian. In the conventional approach, gravitational theories break this symmetry exhibited by all other physical systems. Restoring this symmetry to gravity and demanding that gravitational field equations should also remain invariant under the addition of a constant to a Lagrangian, \textit{leads to} the interpretation of gravity as the thermodynamic limit of the kinetic theory of atoms of space. This approach selects, in a very natural fashion,  Einstein's general relativity in $d=4$. Developing this paradigm at a deeper level, one can obtain the distribution function for the  atoms of space and connect it up with the thermodynamic description of spacetime. This extension relies on a curious fact that the quantum spacetime endows each event with a finite area but zero volume. This approach allows us determine the numerical value of the \cc\ and suggests a new perspective on cosmology.
\end{abstract}

\section{The importance of being hot}

A crucial fact about normal matter, say, a glass of water, which is almost never stressed in textbooks is the following: \textit{You could have figured out that water must be made of discrete atoms without ever probing it at scales comparable to atomic dimensions!} All you need to realize is that water can be heated and hence must have an internal mechanism to store the energy which you supply to it. This is the breakthrough in the understanding of the nature of heat and temperature \cite{tpr:heat} which came with the work of Boltzmann, who essentially said: ``If you can heat it, it has microscopic degrees of freedom''. This profound insight   underscores the following fact: The existence of microscopic degrees of freedom leaves a clear signature \textit{even at the largest macroscopic scales} in the form of temperature and heat. 
One can even count the number of atoms\footnote{Throughout this article I will use the word `atoms' when I mean `microscopic degrees of freedom' or `number of relevant microstates'; they usually differ by an unimportant numerical factor.}, using purely macroscopic variables through the relation
\begin{equation}
Nk_B=R=\frac{E}{(1/2)T} 
\label{bolteq}
\end{equation} 
Again, standard textbooks do not stress the beauty of this result. The variables
 in the right hand side, $E$ and $T$, have valid interpretations in the continuum (thermodynamic) limit, but $N$ in the left hand 
side has no meaning in the same limit. The $N$ actually counts the  number of atoms in the system, the very existence of which is not recognized by continuum thermodynamics!
So you don't need the technology capable of probing matter at angstrom scales in order to figure out that matter is actually made of atoms. The mere fact that matter can be hot, is enough --- if you are as clever as Boltzmann. 

The key new variable which distinguishes thermodynamics from point mechanics is the heat content $TS$ of matter which is the difference $F-E$ between the free energy and internal energy of the system. Expressed in terms of densities, the Gibbs-Duhem relation
(for systems with zero chemical potential) tells us that $Ts = P +\rho$ where $s$ is the entropy density, $\rho$ is the energy density and $P$ is the pressure. The heat density is something uniquely thermodynamic and has no direct analog in point mechanics. More is \textit{indeed} different. 

Let us now proceed from normal matter to the fabric of spacetime. Work done in the last several decades  \cite{A1,A3,A5,A7,A8} shows that even spacetime, like matter, can possess a heat density. As I will soon describe, it is possible to associate a temperature and entropy density with every event in spacetime just as you could have done it to a glass of water. 
On the other hand, one traditionally described the dynamics of spacetime through some  field equation for gravity because Einstein told us that gravity is nothing but the curvature of spacetime. If we take both these results seriously,  we are led to the following conclusions:

\begin{itemize}
 \item The Boltzmann principle  tells us that if spacetime can be hot, it must have microstructure. What is more, we should be able to count the atoms of spacetime without having the technology to do Planck scale experiments just as Boltzmann guessed the existence of atoms of matter without doing angstrom scale experiments. We would expect a relation like \eq{bolteq} to exist for the spacetime.

\item 
If the spacetime is like a fluid made of atoms, the gravitational field equations must have the same status as the equations describing, say, the flow of water. Therefore, we should be able to derive them from a purely thermodynamic variational principle.  Further, the equation itself should allow  a reinterpretation in a purely thermodynamic language rather than in the conventional geometrical language. Consequently, we would expect several variables, which are usually considered geometrical, to have an underlying thermodynamic interpretation. 

 \item The discreteness of matter can be taken into account in the kinetic theory by introducing a distribution function $f(x^i, p_i)$ such that $dN = f(x^i, p_i) d^3x d^3p$ counts the number of atoms in a phase volume. Such a description recognizes discreteness but yet works at scales such that the volume $d^3x$ is large enough   to, say, contain sufficient number of atoms. We should be able to develop a similar concept for spacetime which recognizes the discreteness at the Planck scale and yet allows the use of continuum mathematics to describe the phenomena. 
\end{itemize}

It turns out that the above goals can indeed be achieved thereby providing a thermodynamic description of (what we call) gravity. Further, such an approach allows us to understand several aspects of conventional gravity at a deeper level and --- most importantly --- provides a novel perspective on cosmology capable of predicting the numerical value of cosmological constant. I will now describe how all these  come about. 

\section{Why are spacetimes hot?}

The temperature of a material object is purely kinematic in the sense that a metal rod and a glass of water --- having completely different structural properties --- can possess the same temperature. Similarly, one can associate  a temperature with an event in  a spacetime  which is completely independent of the field equations of gravity which determined the structure of that spacetime. Let me briefly describe how one arrives at this concept just from the kinematics of spacetime. 

Principle of equivalence tells us that (i) the gravitational field is described by the metric of a curved spacetime and  (ii) one  can  determine the influence of gravity on all other systems by a judicious application of the laws of special relativity in the freely falling frame (FFF), around any event. This influence can be completely encoded in the equation $\nabla_a T^a_b =0$ where $T^a_b$ is the symmetric energy momentum tensor of the matter and the $\nabla_a$ depends on the background geometry describing a gravitational field. Applying this to the electromagnetic field, one finds that gravity affects the propagation of light rays and thus the causal structure. 
In particular, it is easy to construct observers (i.e., timelike congruences) in \textit{any} spacetime such that part of the spacetime will be inaccessible to them.  A generic example of such an observer is provided by the local Rindler observers \cite{A35} constructed as follows: Start with the FFF around any event $\mathcal{P}$, with coordinates $(T, \mathbf{X})$ and boost to a local Rindler frame  (LRF) with coordinates $(t,\mathbf{x})$ constructed using some acceleration $a$, through the  transformations: $X=x\cosh (at), T=x\sinh (at)$. There will be a null surface passing though $\mathcal{P}$ which will be the $X=T$ surface in the FFF; this null surface will now act as a patch of horizon to the $x=$ constant Rindler observers.

This LRF leads to the most beautiful result \cite{A5} we know of that arises on combining the principles of general relativity and quantum field theory: The local vacuum state, of the freely falling observers around an event, will appear as a thermal state to the local Rindler observer with the temperature:  
\begin{equation}
 k_BT = \left(\frac{\hbar}{c}\right) \left(\frac{a}{2\pi}\right)
\end{equation}  
where $a$ is the acceleration of the local Rindler observer which can be related to other geometrical variables of the spacetime if required. 
This  temperature tells us that around \textit{any} event, in \textit{any} spacetime, there exist  observers who will perceive the spacetime as hot. 
These local Rindler observers will also notice that  matter takes a very long time to cross the local Rindler horizon thereby allowing for  thermalization to take place.  Since the local Rindler observer attributes a temperature $T$ to the horizon, she will interpret the energy   associated with the matter that crosses the null surface (asymptotically), as  some amount of  energy $\Delta E$ being dumped on a \textit{hot} surface, thereby contributing a \textit{heat} content $\Delta Q_m=\Delta E$.  One can show that the resulting  heat density (energy per unit area per unit  time) of the null surface,  contributed by matter dumped a local Rindler horizon,  as  interpreted by the local Rindler observer,  is given by
\begin{equation}
 \mathcal{H}_m[\ell_a]\equiv \frac{dQ_{m}}{\sqrt{\gamma}d^{2}xd\lambda}=T_{ab} \ell^a\ell^b
\label{hmatter}  
\end{equation}
The  heat transfered by matter is obtained by integrating   $\mathcal{H}_m$ with the integration measure $d\Sigma\equiv\sqrt{\gamma}d^{2}xd\lambda$ over the null surface generated by the null congruence $\ell ^{a}$, parametrized by $\lambda$. (The factor $\sqrt{\gamma}d^{2}x$ is the transverse area element of the $\lambda=$ constant cross-section of the null surface.)
There are two features which are noteworthy regarding $\mathcal{H}_m$.
\begin{itemize}
 \item If we add a constant to the matter Lagrangian (i.e., $L_{m} \to L_{m} + $ constant, the $T^a_{b}$ changes by $T^a_b \to T^a_b + $ (constant) $\delta^a_b$. The  $\mathcal{H}_m$, defined by Eq.~(\ref{hmatter}) remains invariant under this transformation. 

\item  The heat density vanishes if $T^a_b \propto \delta^a_b$. So the cosmological constant has \textit{zero heat density} though it has non-zero energy density. 
In fact, for an ideal, comoving fluid,  $T_{ab} \ell^a\ell^b = (\rho + P)$ and hence the heat density vanishes only for the \cc\ with equation of state $\rho=-P$.
\end{itemize}

Thus the kinematics of spacetime allows us to associate an (observer dependent) temperature with every event in spacetime and a heat density contributed by matter with every null surface. Our next job is to develop a thermodynamic variational principle to obtain the dynamics of the spacetime.

\section{The guiding principle for gravitational dynamics}

Recall that the equations of motion for matter remain invariant when we add a constant to the  Lagrangian. It seems reasonable to postulate that the gravity should not break this symmetry which is  present in the matter sector. Since $T_{ab}$ is the most natural source for gravity (as can be argued from the principle of equivalence and considerations of the Newtonian limit),  this leads to the demand:
\begin{itemize}
 \item[$\blacktriangleright$] The extremum principle which determines the dynamics of spacetime must remain invariant under the shift $T^a_b \to T^a_b + $ (constant) $\delta^a_b$.
\end{itemize}
It can be easily proved \cite{C7} that this  principle  rules out the possibility of varying the metric tensor $g_{ab}$ (in an unrestricted manner) in a covariant, local, action principle to obtain the field equations!   Therefore, our variational principle  cannot use $g_{ab}$ as the dynamical variables and we need to introduce some other auxiliary variables. 
Further, in the traditional approach, $T_{ab}$ arises as the source
when
we vary the metric in the \textit{matter} Lagrangian. But since we are not varying $g_{ab}$, but still want $T_{ab}$ to be the source, we need to explicitly include $T_{ab}$  in the variational principle. So  the variational principle has to \textit{depend} on $T_{ab}$ and yet \textit{be invariant} under $T^a_b \to T^a_b + $ (constant) $\delta^a_b$. The simplest choice (involving the least number of auxiliary degrees of freedom) will be to demand that  the variational principle has the form:
\begin{equation}
Q_{\rm tot}\equiv \int  d\Sigma (\mathcal{H}_m+\mathcal{H}_g); \qquad \mathcal{H}_m[n_a] \equiv T_{ab} n^a n^b
\label{Qtot}
\end{equation}
where the null vector $n_a$ acts as the auxiliary variable. Since $ Q_{\rm tot}$
depends (linearly) on $T_{ab}$ only through the heat density $\mathcal{H}_m[n_a]$ in  \eq{hmatter}, it is obviously invariant under the shift $T^a_b \to T^a_b + $ (constant) $\delta^a_b$.  
The $\mathcal{H}_g$  is the corresponding  contribution from gravity which
is yet to be determined. This approach introduces an arbitrary null vector $n_a$ into the variational principle  which, at this stage, is just an auxiliary field. But since no null vector is special,   the extremum condition should hold for all $n_a$,  leading to a constraint on the background metric $g_{ab}$ thereby determining the  dynamics of spacetime. 

Obviously, the form of the gravitational heat density $\mathcal{H}_g$ determines the spacetime structure, just as the form of entropy functional determines the structure of a material body. A natural choice \cite{A14,A16} for $\mathcal{H}_g$, which is quadratic in $\nabla n$ will have the form:
\begin{equation}
\mathcal{H}_g= -\left(\frac{1}{16\pi L_P^2}\right) (4P^{ab}_{cd}\nabla_an^c\nabla_bn^d)
\label{hgrav}
\end{equation}
where $P^{ab}_{cd}$ is a dimensionless tensor to be determined and $L_P^2$ is an arbitrary constant, with  the dimensions of area. (This expression, by itself, may not look thermodynamical but it is indeed the heat density of gravity --- which should be obvious from the fact that we are adding it to the heat density of matter. This will become clearer later on, see \eq{Qtot1}.)
We require that the condition, $\delta Q_{tot}/\delta n_a=0$ for all null vectors $n_a$ at a given event, should constrain the background geometry. This requirement  leads to the expression:
\begin{equation}\label{Paper06_SecLL_07}
P^{ab}_{cd} \propto  \delta ^{aba_{2}b_{2}\ldots a_{m}b_{m}}_{cdc_{2}d_{2}\ldots c_{m}d_{m}}
R^{c_{2}d_{2}}_{a_{2}b_{2}}\ldots R^{c_{m}d_{m}}_{a_{m}b_{m}}
\end{equation}
where $\delta ^{aba_{2}b_{2}\ldots a_{m}b_{m}}_{cdc_{2}d_{2}\ldots c_{m}d_{m}}$ is the totally antisymmetric $m$-dimensional determinant tensor. The resulting field equation is identical to that of  (what is known as) the \LL\ model \cite{A13,A14,A16} with the \cc\ appearing as an integration constant.
(These models have the interesting --- and unique --- feature that,  the  field equations are  second degree in $g_{ab}$!) The `entropy tensor' $P^{ab}_{cd}$ determines the entropy of horizons  in the resulting theory  through the expression \cite{A8,A13}:
\begin{equation}
s=-\frac{1}{8} \sqrt{\gamma} P^{abcd}\epsilon _{ab}\epsilon _{cd}
\label{entrophys}
\end{equation}
(where $\epsilon_{ab}$ is the binormal to the horizon surface)
One can show that the on-shell value of $Q_{\rm tot}$ is indeed (the difference in) the corresponding heat density of the theory:
\begin{equation}
 Q_{\rm tot}= \int d^2x (T_{\rm loc}\, s)\big|^{\lambda_2}_{\lambda_2}
 \label{Qtot1}
\end{equation} 
(This result also confirms that the $\mathcal{H}_g$ --- which is added to the heat density of matter --- can indeed be interpreted as the heat density of gravity.)
Thus, the specification of horizon entropy specifies the $P^{ab}_{cd}$ and selects the corresponding \LL\ model. The temperature of the spacetime, as we saw before, is purely kinematic, but specifying the form of horizon entropy in \eq{entrophys}, specifies the dynamics of the theory. This is precisely what we expect in the thermodynamic description of a system.

In $d=4$ dimensions, $P^{ab}_{cd}$ reduces to 
$
P^{ab}_{cd}=(1/2)(\delta^a_c\delta^b_d-\delta^b_c\delta^a_d)
$.
The resulting equation for the background spacetime is identical to Einstein's equation:  
\begin{equation}
G^a_b=(8\pi L_P^2)T^a_b+\Lambda \delta^a_b 
\label{efe}                                         
\end{equation} 
with an undetermined cosmological constant.
 By the very construction, the cosmological constant (for which $T_{ab}^{(\Lambda)}n^an^b=0$) cannot appear in the extremum principle; but since the theory is invariant under the shift $T^a_b\to T^a_b+(\mathrm{constant})\delta^a_b$, it arises as an integration constant. (So we need a \textit{further} principle to fix its value once and for all. I will come back to this issue later on.)

Equation (\ref{efe}) is written in the standard form found in the  text books.
It is, however, quite incongruous to claim that gravity is thermodynamics and then write the field equations in terms of conventional geometrical language! I will briefly describe how one can rewrite the same equation in a purely thermodynamic language and also use it to count the density of atoms of spacetime in exactly the same manner as we could use \eq{bolteq} to count the atoms of matter. 

To begin with, let us consider any \textit{static} spacetime foliated by a series of spacelike hypersurfaces. Let $\mathcal{V}$ be a 3-dimensional region in a spacelike hypersurface with a 2-dimensional boundary $\partial \mathcal{V}$, which we could choose to be an equipotential surface (corresponding to constant lapse function).  We can then show that \cite{A17,A18}
the gravitating (Komar) energy $E_{\rm Komar}$ contained in $\mathcal{V}$ is equal to    the equipartition heat energy of the surface $\partial\mathcal{V}$ if we associate $dN=dA/L_P^2$ degrees of freedom with each area element $dA$. That is, we can show:
\begin{equation}
 E_{\rm Komar}=\int \frac{\sqrt{\gamma} \, d^2x}{L_P^2} \left( \frac{1}{2}k_B T_{\rm loc}\right) 
\equiv \frac{1}{2} N_{\rm sur} (k_BT_{\rm avg})
\label{hequi1}
\end{equation} 
where $T_{\rm avg}$ is the average temperature of $\partial \mathcal{V}$ and $N_{sur}=A_{sur}/L_P^2$. 
So we can actually count the microscopic degrees of freedom through an equipartition law which --- since it relates bulk and boundary energies --- could be called holographic equipartition. (One can rescale $(1/2)k_BT\to(\nu/2)k_BT,N_{sur}\to A_{sur}/\nu L_P^2$ without changing the result; we have chosen $\nu=1$.)

We can do  better. Consider the most general spacetime rather than static spacetimes.  In this case, we can associate with the bulk energy $E_{\rm Komar}$ the number  $N_{\rm bulk}$,  defined as the number of degrees of freedom in $\mathcal{V}$ \textit{if}  $E_{\rm Komar}$  is at equipartition at the temperature  $T_{\rm avg}$. That is:
\begin{align}\label{Papper06_NewFin01}
 N_{\rm bulk}\equiv \frac{|E_{\rm Komar}|}{(1/2)T_{\rm avg}}
\end{align}
It then turns out that  \cite{A19}  the time evolution of the spacetime geometry in $\mathcal{V}$ is driven by the 
difference between the 
 bulk and boundary degrees of freedom. Specifically:
\begin{align}\label{Paper06_NewFin02}
\frac{1}{8\pi}\int d^{3}x\sqrt{h}u_a g^{ij}\pounds_\xi N^a_{ij}  =\frac{1}{2}T_{\rm avg}\left(N_{\rm sur}-N_{\rm bulk}\right)
\end{align}
where $N^{c}_{ab}\equiv-\Gamma ^{c}_{ab}+ (1/2)(\delta ^{c}_{a}\Gamma ^{d}_{db}+\delta ^{c}_{b}\Gamma ^{d}_{ad})$,\ 
 $\xi_a\equiv Nu_a$ is the time evolution vector, where $u_a$ is the velocity of the observers moving normal to the foliation.
A simple corollary is that  all static \cite{A17,A18} spacetimes maintain holographic equipartition in terms of  the number of degrees of freedom in the bulk and boundary:
\begin{equation}
 N_{\rm sur}=N_{\rm bulk}
 \label{nsureqn}
\end{equation} 
which, of course, is a nicer restatement of \eq{hequi1}.

The role of Planck constant $\hbar$ in this approach is worth emphasizing. Relativity brings in the speed of light $c$, the Davies-Unruh temperature brings in $\hbar$ and the expression for heat density $\mathcal{H}_g$ introduces the quantum of area $L_P^2$. When we take the Newtonian limit of the gravitational field equations, we will end up getting the gravitational force to be:
\begin{equation}
 F=\left(\frac{c^3 L_P^2}{\hbar}\right) \left(\frac{m_1m_2}{r^2}\right)
 \label{nqg}
\end{equation} 
We should resist the temptation to call the combination $(c^3L_P^2/\hbar)$ as $G$ which is independent of $\hbar$.
Equation~(\ref{nqg}) tells us that the $\hbar \to 0$ limit does not exist: \textit{Gravity is quantum mechanical at all scales!} Just as matter is quantum mechanical at all scales --- the individual atoms will collapse $\hbar \to 0$ limit --- the spacetime and gravity are also quantum mechanical at all scales. The Planck constant plays a more crucial role in this approach than in the usual paradigm.

\section{Distribution function for the atoms of space}

The above discussion highlights the clear analogy between, say, a fluid  and the spacetime from a thermodynamic perspective. The equipartition laws in \eq{bolteq} and  \eq{hequi1}, in particular, allow us to count the number density of atoms in either of these systems. The next logical step will be to take these ideas one level deeper and obtain the gravitational  heat density  $\mathcal{H}_g$ from microscopic considerations. 

To do this, 
 we need to take into account the discreteness of spacetime, arising from the quantum of area $L_P^2$, without losing the privilege of using continuum mathematics in our description. In the case of normal fluid, the use of a distribution function allows us to reconcile these two mutually contradictory requirements. When you say $dN = f(x^i, p_i) d^3x d^3p$ counts the number of atoms around an event $x^i$ (with momentum $p_i$) in a small phase volume, you are assuming that $d^3x $ is small enough to be considered infinitesimal and yet big enough to contain sufficiently large number of atoms;
 by the very process of counting, $f(x^i, p_i)$ incorporates discreteness while allowing the use of continuum mathematics. Proceeding by analogy, we are looking for a distribution function $f(x^i, n_j)$ which could count the number of atoms of space at an event $x^i$ with an additional dynamical variable $n_j$. One might guess that $n_j$ could possibly be related to the null vector  which occurs in the gravitational heat density $\mathcal{H}_g$ 
but this remains to be obtained from the microscopic analysis.

Since  the distribution function has to be a primitive construct in the spacetime, 
it seems natural to assume that \textit{the number of atoms of spacetime will be proportional either to the volume measure or the area measure associated with a given event}. In the continuum description of spacetime, an event has zero area or volume associated with it and hence we cannot hope to obtain a $f(x^i,n_j)$ from such a construction. This is, however, understandable. Unless we incorporate the quantum of area into the description of spacetime, one cannot hope to get a sensible distribution function for the atoms of spacetime. So, our strategy will be to incorporate the zero-point area $L_P^2$ into the fundamental description of spacetime in a suitable manner, define appropriate area and volume measures in such a quantum corrected spacetime and extract the distribution function from these primitive constructs. Of course, there is no guarantee that the quantum spacetime will endow an event with a non-zero area (or volume) but, incredibly enough, it does.
I will now describe this procedure in some detail.

Before we start, it is convenient to re-write $\mathcal{H}_g$ in an equivalent dimensionless form. Using the fact that  
$\mathcal{H}_g \propto 2P^{ab}_{cd}\nabla_an^c\nabla_bn^d$ and $\mathcal{K}_g \propto R_{ab}n^an^b$ differ by an ignorable total divergence  \cite{A19} , we could as well use
the $\mathcal{K}_g$ instead of  $\mathcal{H}_g$ in our variational principle. Introducing the appropriate numerical factors, it is convenient to work with the dimensionless combination 
\begin{equation}
\frac{d(\bar Q_g/E_P)}{d(\sqrt{\gamma}d^2xd\lambda/L_P^3)}\equiv\mathcal{K}_{g} \equiv - \frac{L_P^2}{8\pi } R_{ab}   n^an^b 
\label{hk1}                                  
\end{equation} 
It is this quantity which we hope to obtain from some primitive construct in the quantum spacetime.

Our first task is to incorporate the zero-point area into the spacetime, which could be done in a model independent manner along the following lines. There is considerable amount of evidence   \cite{D2}  to suggest that a primary effect of quantum gravity is to modify the geodesic interval $\sigma^2(x,x')$ in a spacetime to another form $S(\sigma^2)$ such that $S(0) \equiv L_0^2$
is a finite constant of the order of $L_P^2$. For illustrative purposes, we will take $S(\sigma^2) = \sigma^2 + L_0^2$ though none of our results depend on this explicit form. One can show that such a modification is equivalent to working with a renormalized spacetime metric (called qmetric) $q_{ab} (x,x'; L_0^2)$ instead of the original classical metric $g_{ab}(x)$. The explicit form of the qmetric is given by
\begin{align}
q_{ab}=Ah_{ab}+ B n_{a}n_{b};\qquad q^{ab}=\frac{1}{A}h^{ab}+\frac{1}{B}n^{a}n^{b}
\label{qab}
\end{align}
where
\begin{align}
B=\frac{\sigma ^{2}}{\sigma ^{2}+L_{0}^{2}};\qquad A=\left(\frac{\Delta}{\Delta _{S}}\right)^{2/D_{1}}\frac{\sigma ^{2}+L_{0}^{2}}{\sigma ^{2}};\qquad n_a=\nabla_a\sigma
\label{defns}
\end{align}
and $\Delta$ is  the Van Vleck determinant related to the geodesic interval $\sigma^2 $ by
\begin{align}
\Delta (x,x')=\frac{1}{2}\frac{1}{\sqrt{g(x)g(x')}}\textrm{det}\left\lbrace \nabla _{a}^{x}\nabla _{b}^{x'}\sigma ^{2}(x,x') \right\rbrace
\label{vv}
\end{align}
The $\Delta_S$ is the corresponding quantity computed with $\sigma^{2}$ replaced by $S(\sigma^{2})$ in \eq{vv}.

The qmetric is a bi-tensor depending on $x$ and $x'$ through $\sigma^2(x,x')$ and is singular everywhere in the spacetime in the limit of $x'\to x$ with finite $L_0$. On the other hand, $q_{ab} \to g_{ab}$ when $L_0\to 0$ at all events.
Given some scalar $\Phi[g_{ab}(x)]$   constructed from the background metric and  its derivatives, 
we can  compute the corresponding 
  (bi)scalar $\Phi[q_{ab}(x,x');L_0^2]$ for the renormalized spacetime by replacing $g_{ab}$ by $q_{ab}$ in $\Phi[g_{ab}(x)]$ and evaluating all the derivatives at $x$, keeping $x'$ fixed. The renormalized value of $\Phi[q_{ab}(x,x');L_0^2]$ is then obtained by taking the limit $x\to x'$ in this expression keeping $L_0^2$ non-zero.  It turns out that many useful scalars like $R$, $K$ etc. remain finite
  \cite{D1,D5,D6}
  and local in this limit even though the qmetric itself is singular when $x\to x'$ with non-zero $L_0^2$. The algebraic reason for this curious fact  \cite{D1} is that the following two limits do not commute:
\begin{equation}
\lim_{L_0^2\to 0}\, \lim_{x\to x'} \Phi[q_{ab}(x,x');L_0^2]\neq  \lim_{x\to x'}\,\lim_{L_0^2\to 0} \Phi[q_{ab}(x,x');L_0^2]
\end{equation}

It is now easy to see how null surfaces and null vectors are singled out in this approach.  In all calculations we will eventually take the limit $\sigma^2 \to 0$ in the Euclidean sector.
 But 
this limit, $\sigma^2 \to 0$, will translate into a null surface in the Minkowski spacetime\footnote{The local Rindler observers who live on the hyperboloid $r^2-t^2=\sigma^2$ see the null cone $r^2-t^2=0$  as the horizon. In the Euclidean sector the hyperboloid becomes the sphere $r^2+t_E^2=\sigma_E^2$ and approaching the Euclidean origin, $\sigma_E\to0$, translates to approaching the light cone in the Minkowski space.} and the normal vector $n_i = \nabla_i\sigma$ (which occurs in the qmetric and all the resulting constructs) will pick out the null vector which is the normal to the null surface. More generally, $\sigma^2 (x,x') \to 0$ selects out events which are connected by a null geodesic and hence $n_a$ will correspond to a null vector in the Minkowski spacetime.
This is how a null vector field $n_i$ is introduced in the description from a microscopic point of view.

With this mathematical structure in place, we can define the volume and area measure of the renormalized spacetime as follows. 
It is convenient to describe the Euclidean background spacetime in synchronous coordinates $(\sigma, \theta_1, \theta_2, \theta_3)$ where $\sigma$ (the geodesic distance from the origin) is the `radial' coordinates and $\theta_i$ are the angular coordinates on the 
equi-geodesic surfaces corresponding to $\sigma =$ constant.
 We next introduce the zero-point-area by constructing the corresponding qmetric. (The equigeodesic surfaces remain equi-geodesic surfaces  in the renormalized spacetime.) Using the qmetric one can then compute the volume measure, $\sqrt{q} d^4 x$, as well as the area measure of the equi-geodesic surfaces, $\sqrt{h} d^3x$. Both $\sqrt{q}$ and $\sqrt{h}$ will be now bi scalars and we define their value at a given event by taking the limit of $x'\to x$ corresponding to $\sigma^2 \to 0$. As mentioned earlier, this will lead to a dependence on a null vector $n_i$ which could be in any direction at the given event (and  is reminiscent of  the momentum variable which occurs in the distribution function of normal matter.)   
Such a computation shows that the volume and area measures behave as follows:\footnote{
These results are somewhat subtle algebraically.  The leading order behaviour of $\sqrt{q}d\sigma\approx\sigma d\sigma$, which makes the volumes scale as $\sigma^2$ (while the area measure is finite) produces  the following result \cite{paperD}: \textit{The effective dimension of the renormalized spacetime reduces to $D=2$ close to Planck scales.} I will not elaborate on this result here.
}
\begin{align}
\sqrt{q}=\sigma \left(\sigma ^{2}+L_{0}^{2}\right)\left[1-\frac{1}{6}\mathcal{E}\left(\sigma ^{2}+L_{0}^{2}\right)\right]\sqrt{h_\Omega}
\label{qfinal}
\end{align}
\begin{align}
\sqrt{h}=\left(\sigma ^{2}+L_{0}^{2}\right)^{3/2}\left[1-\frac{1}{6}\mathcal{E}\left(\sigma ^{2}+L_{0}^{2}\right)\right]\sqrt{h_\Omega}
\label{hfinal}
\end{align}
where $\mathcal{E} \equiv R_{ab}n^an^b$. 
When $L_{0}^{2}\to0$ we recover the result in classical differential geometry known to Gauss, as we should. But our interest is in the limit $\sigma^2\to0$ at finite $L_0$.
Something remarkable  happens when we do this. The volume measure $\sqrt{q}$ vanishes but the area measure $\sqrt{h}$ has a non-zero limit given by:
\begin{align}
\sqrt{h}= L_{0}^{3}\left[1-\frac{1}{6}\mathcal{E}L_{0}^{2}\right]\sqrt{h_\Omega}
\label{hlimit}
\end{align}
The  the renormalized spacetime
attributes to every point in the spacetime a finite area measure but a zero volume measure! 
Since $L_0^3\sqrt{h_\Omega}$ is the volume measure of the $\sigma=L_0$ surface, the dimensionless density of the atoms of spacetime, contributing to the gravitational heat is given by:
\begin{equation}
f(x^i,n_a)\equiv \frac{\sqrt{h}}{L_0^3\sqrt{h_\Omega}} =1-\frac{1}{6}\mathcal{E}L_{0}^{2}
=1-\frac{1}{6} L_{0}^{2} R_{ab}n^an^b
\label{denast}
\end{equation} 
This matches with what we need if we take $L_0^2=(3/4\pi) L_P^2$.
Briefly stated, quantum gravity endows each event in spacetime with a finite area but zero volume. It is this area measure which we compute to obtain a natural estimate for $f(x^i,n_a)$.
In the macroscopic limit, 
the contribution to the  gravitational heat in any volume is obtained by integrating $f(x^i,n_j)$ over the volume. So the expression for the heating rate, in dimensionless form  is  given by:
\begin{equation}
L_P^2\frac{dQ_{g}}{d\lambda}=\int\frac{\sqrt{\gamma}d^2x}{L_P^2}f(x^i,n_j) =
\int\frac{\sqrt{\gamma}d^2x}{L_P^2}\left[1-\frac{1}{8\pi}L_{P}^{2}(R_{ab}n^an^b)\right]
\label{corres}
\end{equation} 
which gives the the correct expression --- with the crucial minus sign --- plus a constant. 
\textit{So one can indeed interpret the gravitational heat density as the area measure of the renormalized spacetime.}

While the second term in \eq{denast} gives what we want for the variational principle, the first term tells us that there is a zero-point contribution to the degrees of freedom in spacetime, which, in dimensionless form, is just unity. Therefore, it makes sense to ascribe $A/L_P^2$ degrees of freedom  to an area $A$, which is consistent with what we saw in the macroscopic description.
We also see that a two sphere of radius $L_P$ has $4\pi L_P^2/L_P^2=4\pi$ degrees of freedom. This was the crucial input which was used in a previous work to determine the numerical value of the \cc\ for our universe. 
Using this result, one can show express the energy density corresponding to the \cc\ in the form \cite{C8,C9}:
\begin{equation}
 \rho_\Lambda  \approx \frac{4}{27} \frac{\rho_{\rm inf}^{3/2}}{\rho_{\rm eq}^{1/2}} \exp (- 36\pi^2)
\label{ll6}
 \end{equation} 
where $\rho_{\rm inf}$ is the energy density during inflation and $\rho_{eq}$ is the energy density at the epoch of matter radiation equality.
From cosmological observations, we find that  $\rho_{eq}^{1/4} = (0.86 \pm 0.09) \text{eV}$; if we take the range of  the inflationary energy scale  as $\rho_{\rm inf}^{1/4} = (1.084-1.241) \times 10^{15}$ GeV, we get $\rho_{\Lambda} L_P^4  = (1.204 - 1.500) \times 10^{-123}$, which is consistent with observations! 

This  novel approach for solving the cosmological constant problem provides a unified view of cosmic evolution, connecting all the three phases through \eq{ll6}; this is to be contrasted with standard cosmology in which the three phases are put together in an unrelated, ad hoc manner.
Further, this approach to the cosmological constant problem \textit{makes a falsifiable prediction}, unlike any other approach I know of. From the observed values of $\rho_\Lambda$ and $\rho_{\rm eq}$ we can constrain the energy scale  of inflation to a very narrow band --- to within a factor of about five, if we consider the ambiguities in re-heating. If future observations show that inflation took place at energy scales outside the band of $(1-5)\times 10^{15}$ GeV, this model for explaining the value of \cc\ is ruled out.

\section{Outlook}

We have completed the program outlined in the introduction using essentially two  ingredients: (a) We postulated that the extremum principle determining spacetime dynamics should be invariant under the shift $T^a_b \to T^a_b + \text{(constant)}\, \delta^a_b$. this allowed us to obtain an expression for gravitational heat density which depended on a null vector that acted as an auxiliary variable.  (b) We introduced the zero-point area into the spacetime by the replacement $\sigma^2 \to \sigma^2 + L_0^2$. The modified spacetime led to an area measure which, in dimensionless form, matched precisely with the gravitational heat density we needed.  We interpreted the microscopic origin in terms of the distribution function for the atoms of spacetime. 

This approach raises several important issues for further investigations and let me mention a  couple of them. 
First, we need to understand precisely \textit{what} is counted by $f(x^i,n_j)$. We called it atoms of space which stands for the microscopic degrees of freedom of quantum space(time) parametrized by a null vector $n_i$. One could equally well have thought of it as related to number of microscopic states available to quantum geometry. This suggests that, in the suitable limit, one can introduce a probability $P(x^i,n_a)$ for $n_a$ at each event $x^i$ and define the partition function:
\begin{equation}
 e^{S(x^i)}\propto \int\mathcal{D}n_i P(x^i,n_a)\exp[\mu L_P^4 T_{ab}n^an^b]
\end{equation} 
where $\mu$ is a numerical factor of order unity. If we take 
\begin{equation}
 P(x^i,n_a)\propto\exp [\mu f(x^i,n_a)] \propto \exp\left( - \frac{\mu L_P^2}{8\pi} R_{ab} n^an^b\right)
 \label{eqnx}
\end{equation} 
then the saddle point evaluation will peak at the geometry determined by Einstein's equation with an arbitrary cosmological constant. (The choice $\mu=1/4$ will allow $P$ to be interpreted as number of microstates.) Alternatively, one can think of $P(x^i,n_a)$ to be such as to give the 
correlator 
$
 \langle n^an^b\rangle \approx(4\pi/\mu L_P^2) R_{ab}^{-1}
 $
which allows us to write the field equations in the form:
 \begin{equation}
  2\mu L_P^4\ \langle \bar T_{ab} n^an^b\rangle\approx 2\mu L_P^4\ \langle \bar T_{ab}\rangle \langle n^an^b\rangle =1
\label{mach}
 \end{equation} 
 The averaging $\langle \cdots \rangle$ now indicates both expectation values for the quantum operator $T_{ab}$ as well as a probabilistic averaging of $n^an^b$. 
Equation~(\ref{mach}) has a Machian flavour. One cannot set $\langle T_{ab}\rangle =0$ and study the resulting spacetime since it will lead to $0=1$!. \textit{Matter and geometry must emerge and co-exist together in a manner we have not yet understood.} There is no such thing as flat spacetime existing in the absence of matter! 

Second, a thermodynamic approach to gravity strongly suggests that cosmology should \textit{not} be treated as a part of general relativity and we should look at cosmic questions  afresh.  The study of thermodynamics, distribution functions for atoms of space etc. pre-supposes some unstated notion of equilibrium at the microscopic scales, which, in turn, will involve certain timescales over which such an equilibrium can be established. For normal systems characterized by timescales much less than the  age of the universe, one could possibly assume that Planck scale physics has established the necessary equilibrium conditions. But such an assumption is likely to break down when we consider the entire universe as a physical system. Instead, one is led to a picture in which larger and larger spatial scales achieve microscopic equilibrium as the cosmic time evolves. In such a scenario, one could even argue that the space as we know itself emerges \cite{tpcosmos} as a condensate of the atoms of space as the cosmic time evolves. The deviations from microscopic equilibrium can then have important implications for the large scale dynamics of the universe, a glimpse of which was seen in the suggested solution to the cosmological constant problem.

\end{document}